\documentclass{iopart}

\usepackage[dvips]{graphicx}
\usepackage{bm}

\def\program#1{{\sc #1}}

\begin{document}
 
\title{A new spectral apparent horizon finder for 3D
numerical relativity}

\author{Lap-Ming Lin$^{1,2,3}$\footnote{Email address: lmlin@phy.cuhk.edu.hk} 
and J\'{e}r\^{o}me Novak$^3$\footnote{Email address: Jerome.Novak@obspm.fr} }
\address{$^1$ Department of Physics and Institute of Theoretical Physics, 
The Chinese University of Hong Kong, Hong Kong, China}
\address{$^2$ Department of Physics, University of Hong Kong, Hong Kong, 
China}
\address{$^3$ Laboratoire de l'Univers et de ses Th\'{e}ories, 
UMR 8102 du CNRS, Observatoire de Paris, F-92195 Meudon Cedex, France}

\date{\today}
 
\begin{abstract} 
  We present a new spectral-method-based algorithm for finding apparent
  horizons in three-dimensional space-like hypersurfaces without symmetries.
  While there are already a wide variety of algorithms for finding apparent
  horizons, our new algorithm does not suffer from the same weakness as
  previous spectral apparent horizon finders: namely the monopolar coefficient
  ($\ell=0$ in terms of the spherical harmonics decomposition) needed to be
  determined by a root-finding procedure. Hence, this leads to a much faster
  and more robust spectral apparent horizon finder. The finder is tested with
  the Kerr-Schild and Brill-Lindquist data.  Our finder is accurate and is as
  efficient as the currently fastest methods developed recently by Schnetter
  (2003 Class. Quantum Grav. {\bf 20}, 4719) and Thornburg (2004 Class. Quantum
  Grav. {\bf 21}, 743). At typical resolutions it takes only 0.5 second
  to find the apparent horizon of a Kerr-Schild black hole with $a=0.9M$ to
  the accuracy $\sim 10^{-5}$ for the fractional error in the horizon's
  location on a 2 GHz processor.

\end{abstract}


%%\maketitle

\def\D{ \overline{D} }

%%%%%%%%%%%%%%%%%%%%%%%%%%%%%%%%%%%%%%%%%%%%%%%%%%%%% 
\section{\bf Introduction} 
\label{sec:intro}
%%%%%%%%%%%%%%%%%%%%%%%%%%%%%%%%%%%%%%%%%%%%%%%%%%%%% 

Apparent horizons play an important role in numerical relativity for 
spacetimes containing black hole(s). Being defined locally in time 
(see section~\ref{sec:define}), the apparent horizon(s) can readily be 
computed from the data on each hypersurface during a numerical evolution 
in ($3+1$) numerical relativity. 
In contrast, the event horizon is a global property and can be determined 
approximately only when the spacetime has essentially settled down to a 
stationary state. Once the spacetime has settled down, the event horizon can 
be found at all previous times by integrating null geodesics backwards in time
(e.g., \cite{hug94,ann95,lib96,die03}). 
As it (if exists) must be inside an event horizon \cite{haw73}, the apparent horizon 
is an important tool to track the location and movement of black hole(s) in 
a numerically generated spacetime. Furthermore, the surface of an apparent horizon 
also provides a natural boundary within which the spacetime region can be excised 
from the computational domain in order to handle the physical singularities inside
a black hole \cite{unr84,tho87,sei92} (also see, e.g., \cite{alc03,cam06,van06,bai06} for black 
hole simulations without excision). 
In the new concept of a `dynamical horizon' (see \cite{ash04,boo05} for reviews), apparent 
horizons are essentially the cross sections of the (three-dimensional space-like) 
dynamical horizon on the hypersurfaces. It has recently been shown in this context that 
the areas of the apparent horizons satisfy a causal evolution equation and give a 
positive bulk viscosity in a viscous fluid analogy \cite{gou06}. This is in contrast to 
the event horizon which yields a noncausal evolution and a negative 
bulk viscosity.

A wide variety of algorithms for finding apparent horizons have been proposed 
in the past decade. We refer the reader to the review article by 
Thornburg \cite{tho05} (and references therein) for details. 
In this paper, we present a new apparent horizon finder which is based on 
spectral methods. While the spectral-method-based algorithm for finding apparent 
horizons was first proposed by Nakamura {\it et al} \cite{nak84} more than 20 years ago, our new approach does not suffer from the same weakness as in the 
Nakamura {\it et al} algorithm: namely the $\ell=0$ coefficient of the 
spherical harmonics decomposition of the apparent horizon's surface needed to be 
determined by a root-finding procedure. Hence, our algorithm leads to a more 
robust and efficient spectral apparent horizon finder.  
We have tested our finder with analytic solutions for single and two black-hole
spacetimes. 
Our finder is as efficient as the currently fastest algorithms 
developed by Schnetter \cite{sch03} and Thornburg \cite{tho04}. 

This paper is organized as follows. In section~\ref{sec:define} we present the
notations and various definitions. In section~\ref{sec:nak} we briefly review
the Nakamura {\it et al.} algorithm; we describe our spectral algorithm and
the numerical procedure in section~\ref{sec:our_code}. Section~\ref{sec:tests}
presents tests with analytic solutions to assess the accuracy, robustness, and
efficiency of our finder. Finally, we summarize our results in
section~\ref{sec:conclude}.  Latin (Greek) indices go from 1 to 3 (0 to 3).

%%%%%%%%%%%%%%%%%%%%%%%%%%%%%%%%%%%%%%%%%%%%%
\section{\bf Notations and definitions}
\label{sec:define}
%%%%%%%%%%%%%%%%%%%%%%%%%%%%%%%%%%%%%%%%%

Given a space-like hypersurface $\Sigma$ with future-pointing 
unit normal $n^{\mu}$, the 3-metric $\gamma_{\mu\nu}$ induced by the 
spacetime metric $g_{\mu\nu}$ onto $\Sigma$ is 
\begin{equation}
\gamma_{\mu\nu} := g_{\mu\nu} + n_{\mu}n_{\nu} . 
\end{equation}
Let $S$ be a closed smooth (two-dimensional) surface embedded in $\Sigma$,
and let $s^{\mu}$ be the outward-pointing unit normal of $S$, which is 
spacelike and also normal to $n^{\mu}$ (i.e., $s_{\mu}s^{\mu}=1$ and 
$s_{\mu}n^{\mu}=0$). The 3-metric $\gamma_{\mu\nu}$ now induces a 
2-metric on $S$:
\begin{equation}
m_{\mu\nu} := \gamma_{\mu\nu} - s_{\mu}s_{\nu} . 
\end{equation}
Let $k^{\mu}$ be the tangents of the outgoing future-pointing null 
geodesic whose projection on $\Sigma$ is orthogonal to $S$. We have 
(up to an overall factor)
\begin{equation}
k^{\mu} = s^{\mu} + n^{\mu} ,
\end{equation}
on the 2-surface $S$. 

The expansion of the outgoing null geodesics is 
\begin{equation}
\Theta = \nabla_{\mu} k^{\mu} ,
\end{equation}
where $\nabla_{\mu}$ is the covariant derivative associated with $g_{\mu\nu}$.
In terms of three-dimensional quantities, on the 2-surface $S$, the expansion 
can be written as (see, e.g., \cite{bau03}) 
\begin{equation}
\Theta = D_i s^i - K + s^i s^j K_{ij} ,
\end{equation}
where $D_i$ is the covariant derivative associated with $\gamma_{ij}$, 
$K_{ij}$ is the extrinsic curvature of $\Sigma$ and $K$ is the trace 
of $K_{ij}$. The expansion can also be written as
\begin{equation}
\Theta = m^{ij} \left( D_i s_j - K_{ij} \right) .
\label{eq:Theta_mij}
\end{equation}
The 2-surface $S$ is called a marginally trapped surface if $\Theta=0$
everywhere on $S$. We shall call here the outermost of such surfaces (which is
a marginally outer trapped surface - MOTS) the apparent horizon.

To parameterize the apparent horizon, we assume that the topology of $S$ is a 
2-sphere, and $S$ is star-shaped around the coordinate origin $r=0$, which means 
that for every point $M$ inside $S$, the straight line connecting the 
origin to $M$ is entirely inside $S$ \cite{bon98}. 
The position of the apparent horizon can then be represented as 
\begin{equation}
F(r,\theta,\varphi) := r - h(\theta,\varphi) = 0 , 
\end{equation}
where $(r,\theta,\varphi)$ are the standard spherical coordinates. The function
$h$ measures the coordinate distance to the horizon's surface in the 
direction $(\theta,\varphi)$. With this parametrization, the unit 
normal $s^i$ is given by 
\begin{equation}
s^i = { D^i F \over \left( \gamma^{ij} D_i F D_j F \right)^{1/2} } 
:= { D^i F \over |D F| } , 
\end{equation}
where $D^i := \gamma^{ij} D_j$. The expansion (equation~(\ref{eq:Theta_mij})) 
becomes
\begin{equation}
\Theta = m^{ij} \left( { D_iD_j F\over |DF| } - K_{ij} \right) ,
\label{eq:Theta}
\end{equation}
where the condition $m^{ij} s_j = 0$ has been used. 

%%%%%%%%%%%%%%%%%%%%%%%%%%%%%%%%%%%%%%%%%
\section{\bf The Nakamura {\it et al} algorithm}
\label{sec:nak}
%%%%%%%%%%%%%%%%%%%%%%%%%%%%%%%%%%%%%%%%%

In this section, we give a brief review of the algorithm adopted by 
Nakamura {\it et al} \cite{nak84} for finding apparent horizon 
based on spectral methods. They expand $h$ in spherical harmonics:
\begin{equation}
h(\theta,\varphi) = \sum_{\ell=0}^{\ell_{\rm max} } \sum_{m=-\ell}^{\ell} 
a_{\ell m} Y_\ell^m (\theta,\varphi) .
\label{eq:h_expand}
\end{equation}
They rewrite the apparent horizon equation $\Theta=0$ 
as\footnote{We follow the notation of Gundlach \cite{gun98}.}
\begin{equation}
\Delta_{\theta\varphi} h = \rho \Theta + \Delta_{\theta\varphi} h , 
\label{eq:NKO}
\end{equation}
where $\Delta_{\theta\varphi}$ is the flat Laplacian operator on a 
2-sphere defined by 
\begin{equation}
\Delta_{\theta\varphi} h := h_{,\theta\theta} + \cot\theta
h_{,\theta} + \sin^{-2}\theta h_{,\varphi\varphi} .
\end{equation}
The positive scalar function $\rho$ is chosen such that the term 
$h_{,\theta\theta}$ cancels on the right-hand side (RHS). Using the fact that 
the $Y_\ell^m$ are an orthogonal set of eigenfunctions of $\Delta_{\theta\varphi}$,
\begin{equation}
\Delta_{\theta\varphi} Y_\ell^m = -\ell\left(\ell+1 \right) Y_\ell^m ,
\end{equation}
we obtain the relation (with $d\Omega = \sin\theta d\theta d\varphi$)
\begin{equation}
-\ell\left( \ell+1 \right) a_{\ell m} = \int_S Y_\ell^{m*} \left( \rho\Theta + 
\Delta_{\theta\varphi} h \right) d\Omega .
\label{eq:old_alm}
\end{equation}
This equation can be used to solve for the coefficients $a_{\ell m}$ via an 
iteration procedure. However, the value of $a_{00}$ has to be determined 
at each iteration step by solving for the root of 
\begin{equation}
\int_{S} Y_{0}^{0*} \left( \rho\Theta + \Delta_{\theta\varphi} 
h \right) d\Omega = 0 .
\label{eq:a00}
\end{equation} 

The main disadvantage of the above scheme is that the coefficient $a_{00}$ 
has to be determined separately by  equation~(\ref{eq:a00}). As pointed out by 
Gundlach \cite{gun98}, solving equation~(\ref{eq:a00}) by any iteration method is as 
computationally expensive as many steps of the main iteration loop. Furthermore, 
equation~(\ref{eq:a00}) may have multiple roots or none. In those cases, each root 
or each minimum (if there is no root) should be investigated separately \cite{kem91}.
This clearly reduces the efficiency of the algorithm significantly.

%%%%%%%%%%%%%%%%%%%%%%%%%%%%%%%%%%%%%%%
\section{\bf Our algorithm}
\label{sec:our_code}
%%%%%%%%%%%%%%%%%%%%%%%%%%%%%%%%%%%%%%

\subsection{Master equation for apparent horizon}

Our spectral-method-based algorithm uses a similar ansatz (\ref{eq:NKO}) as 
Nakamura {\it et al} \cite{nak84}. 
The main difference is that we do not need to determine $a_{00}$ separately. 
Hence, this leads to a more robust and efficient apparent horizon finder based 
solely on the spectral method\footnote{See \cite{gun98} for Gundlach's ``fast flow''
algorithm which combines the spectral algorithm of Nakamura {\it et al} and 
the so-called curvature flow method (see also \cite{tho05} for discussion).}. 

To begin, we first introduce a flat metric $f_{ij}$ on the hypersurface $\Sigma$. 
The components of the flat metric with respect to the spherical coordinates 
$(r,\theta,\varphi)$, and the associated natural basis 
$({\partial\over\partial r}, {\partial\over \partial \theta},
{\partial\over\partial \varphi})$, are $f_{ij} = {\rm diag}(1,r^2,r^2 \sin\theta)$. 
Let $\D_i$ be the covariant derivative associated with $f_{ij}$. 
The expansion function $\Theta$ (equation~(\ref{eq:Theta}))
can now be written as 
\begin{equation}
\Theta = \left( \gamma^{ij} - s^i s^j \right)
\left[ |DF|^{-1} \left( \D_i\D_j F - \Delta^m_{\ ij} \D_m F \right) 
- K_{ij} \right] , 
\label{eq:Theta_1}
\end{equation}
where the tensor field $\Delta^m_{\ ij}$ is defined by 
\begin{equation}
\Delta^m_{\ ij} :=
{1\over 2} \gamma^{mn}\left( \D_i \gamma_{jn} + \D_j \gamma_{in}
- \D_n \gamma_{ij} \right) .
\end{equation}
We have also used the following relation between the 
two covariant derivatives $D_i$ and $\D_i$:
\begin{equation}
D_i V_j = \D_i V_j - \Delta^m_{\ ij} V_m ,
\end{equation}
where $V^j$ is an arbitrary 3-vector on $\Sigma_t$. 

Motivated by the recently proposed fully constrained-evolution scheme for 
numerical relativity \cite{bon04}, we define a conformal factor $\Psi$ by 
\begin{equation}
\Psi := \left( { {\rm det} \gamma_{ij}\over {\rm det} f_{ij} } 
\right)^{1/12} , 
\label{eq:Psi}
\end{equation}
and also a tensor field $h^{ij}$ by 
\begin{equation}
\gamma^{ij} = \Psi^{-4} \left( f^{ij} + h^{ij} \right) .
\label{eq:hij}
\end{equation}
We also expand all tensor fields onto the following spherical basis:
\begin{equation}
{\bf{e}}_{\hat{r}} := {\partial\over \partial r} , \ \ 
{\bf{e}}_{\hat{\theta}} := {1\over r}{\partial\over \partial\theta} , \ \
{\bf{e}}_{\hat{\varphi}} := {1\over r \sin\theta} {\partial\over \partial 
\varphi } . 
\end{equation}
This basis is orthonormal with respect to the flat metric: 
$f_{\hat{i}\hat{j}} = {\rm diag}(1,1,1)$. Here and afterwards, we denote 
the tensor indices associated with this basis with a hat. 
The expansion function now becomes 
\begin{eqnarray}
 \Theta &=& \Psi^{-4} |DF|^{-1} f^{\hat{i}\hat{j}} 
\D_{\hat{i}} \D_{\hat{j}} F + \left( \Psi^{-4} h^{ \hat{i}\hat{j} }
- s^{\hat{i}} s^{\hat{j}} \right) |DF|^{-1} \D_{\hat{i}}\D_{\hat{j}} F \cr
&& \cr
&& -\left(\gamma^{\hat{i}\hat{j}} -s^{\hat{i}} s^{\hat{j}}\right)
\left( |DF|^{-1} \Delta^{\hat m}_{\ \hat{i}\hat{j}} \D_{\hat{m}} F 
+ K_{\hat{i}\hat{j}} \right) .
\label{eq:Theta_2}
\end{eqnarray}
Now let us consider the first term on the RHS of this equation:
\begin{eqnarray}
\fl \Psi^{-4}|DF|^{-1} f^{\hat{i}\hat{j}} \D_{\hat{i}} \D_{\hat{j}} F 
&=& \Psi^{-4}|DF|^{-1}\left( \D_{\hat{r}}\D_{\hat{r}}F 
+ \D_{\hat{\theta}}\D_{\hat{\theta}}F
+ \D_{\hat{\varphi}}\D_{\hat{\varphi}}F \right) \cr
&& \cr
&=& {-1\over \Psi^4 |DF| r^2 }\left( h_{,\theta\theta}
+\cot\theta h_{,\theta} + \sin^{-2}\theta h_{,\varphi\varphi} - 2r \right) \cr
&& \cr
&=& {-1\over \Psi^4 |DF| h^2 } \left( \Delta_{\theta\varphi} h - 2 h \right) ,
\label{eq:Theta_1stterm}
\end{eqnarray}
where we have set $r=h(\theta,\varphi)$ for the apparent
horizon in the last equality. 
In equation~(\ref{eq:Theta_1stterm}), we have used the following relation 
for the components of the covariant derivative $\D_{\hat j}$ of a 3-vector 
$V^{\hat i}$ in the orthonormal basis $\{{\bf e}_{\hat i}\}$:
\begin{equation}
\D_{\hat j} V_{\hat i} = e_{\hat j}^{\ k} {\partial \over \partial x^k}
V_{\hat i} - \hat \Gamma^{\hat k}_{\ \hat i \hat j} V_{\hat k} ,
\end{equation}
where $e_{\hat j}^{\ k} := {\rm diag}[1,1/r,1/(r\sin\theta)]$. 
The $\hat{\Gamma}^{\hat{k}}_{\ \hat{i}\hat{j}}$ are the connection coefficients of 
$\D_{\hat k}$ associated with $\{ {\bf e}_{\hat i} \}$. The non-vanishing components 
are
\begin{equation}
\fl 
\hat{\Gamma}^{\hat r}_{\ \hat\theta \hat\theta} = 
- \hat{\Gamma}^{\hat\theta}_{\ \hat r\hat\theta} = - {1\over r} , \ \ \  
\hat{\Gamma}^{\hat r}_{\ \hat\varphi \hat\varphi} =
- \hat{\Gamma}^{\hat \varphi}_{\ \hat r\hat \varphi} = - {1\over r} , \ \ \  
\hat{\Gamma}^{\hat\theta}_{\ \hat\varphi \hat\varphi} = 
- \hat{\Gamma}^{\hat \varphi}_{\ \hat\theta \hat\varphi} = {-1\over r \tan\theta} . 
\end{equation}

Equations~(\ref{eq:Theta_2}) and (\ref{eq:Theta_1stterm}) suggest that, instead of 
the ansatz~(\ref{eq:NKO}) as taken by Nakamura {\it et al}, it is more appropriate to 
rewrite the apparent horizon 
equation $\Theta=0$ as
\begin{equation}
\Delta_{\theta\varphi} h - 2h = \lambda \Theta + \Delta_{\theta\varphi} h - 2h ,
\label{eq:ansatz}
\end{equation}
where the scalar function $\lambda$ is chosen to be $\lambda=\Psi^4 |DF| h^2$ such 
that the combination $\Delta_{\theta\varphi}h-2h$ cancels on the RHS of this equation.  
Hence, the master equation that we solve in our algorithm is 
\begin{eqnarray}
\Delta_{\theta\varphi} h - 2h &=& \Psi^4 |DF| h^2  \left[
\left( \Psi^{-4} h^{ \hat{i}\hat{j} }
- s^{\hat{i}} s^{\hat{j}} \right) |DF|^{-1} \D_{\hat{i}}\D_{\hat{j}} F 
\right. \cr
&& \cr
&& \left. -\left(\gamma^{\hat{i}\hat{j}} -s^{\hat{i}} s^{\hat{j}}\right)
\left( |DF|^{-1} \Delta^{\hat m}_{\ \hat{i}\hat{j} } \D_{\hat{m}} F 
+ K_{\hat{i}\hat{j}} \right) \right] .
\label{eq:master}
\end{eqnarray}
The expansion coefficients are now determined by solving the 
following equation iteratively:  
\begin{equation}
a_{\ell m} = {-1\over \ell\left(\ell+1\right) + 2} \int_S Y_\ell^{m*} \left( \lambda
\Theta + \Delta_{\theta\varphi} h - 2 h \right) d\Omega .
\label{eq:new_alm}
\end{equation} 
This equation applies for all $\ell\geq 0$, and hence $a_{00}$ is not treated 
specially. 
In \cite{shi97}, Shibata developed an apparent horizon finder based essentially on the same form of equation~(\ref{eq:ansatz}). However, he solved the equation using the finite-differencing method without pointing out the key advantage that, if solved by the spectral method, the coefficient $a_{00}$ (as determined by our equation~(\ref{eq:new_alm})
and his equation (1.3) in \cite{shi97}) does not needed to be solved by the root-finding 
procedure. In this work, we solve the algorithm for the first time with spectral method.  
The difference between equations~(\ref{eq:new_alm}) and (\ref{eq:old_alm})
leads to a dramatic improvement in the efficiency and robustness of spectral-method-based algorithms for finding apparent horizons.

\subsection{Numerical procedure}
\label{sec:numerical}

For given 3-metric $\gamma_{ij}$ and extrinsic curvature $K_{ij}$ on a
hypersurface $\Sigma$, equation~(\ref{eq:master}) represents a nonlinear 
elliptic equation for the function $h$. We solve this equation iteratively 
by considering the RHS of the equation as a source term for the linear 
operator $\Delta_{\theta\varphi}-2$ acting on $h$. We use a multidomain 
spectral method to solve the elliptic equation \cite{bon98,bon99}. 
The code is constructed upon the C++ library LORENE \cite{lorene}, and is publicly 
available. 

The numerical iteration procedure is briefly described here.
Assume that the data $(\gamma_{\hat i\hat j},K_{\hat i\hat j})$ are given on $\Sigma$. The conformal 
factor $\Psi$ and the tensor field $h^{\hat i\hat j}$ are then calculated by equations~(\ref{eq:Psi})
and (\ref{eq:hij}), respectively. 
Assume that an initial guess for the function $h(\theta,\varphi)$ is chosen (equivalently for the spectral coefficients $a_{\ell m}$).
The iteration processes are as follows. 

\begin{enumerate}
\item At the $n$th iteration step, the function $h^{(n)}$ is determined by the 
coefficients $a_{\ell m}^{(n)}$ (with the superscript $(n)$ labels the iteration steps).
The level-set function $F$ and the unit normal vector $s^i$ are then obtained from 
$h^{(n)}$ (see section~\ref{sec:define}). 
\item The spectral coefficients at the next iteration step are calculated by 
equation~(\ref{eq:new_alm}):
{\begin{equation}
a_{\ell m}^{(n+1)} = {-1\over \ell(\ell+1) +2 } \int_S Y_\ell^{m*} S^{(n)} d\Omega , 
\end{equation} }
where $S^{(n)}$ represents the RHS of equation~(\ref{eq:master}) evaluated from 
$a_{\ell m}^{(n)}$. 
The new function $h^{(n+1)}$ is then obtained from $a_{\ell m}^{(n+1)}$ 
by equation~(\ref{eq:h_expand}). 
\item The difference between $h^{(n+1)}$ and $h^{(n)}$ is calculated. The iteration
procedure continues until the maximum value of the difference throughout the whole 
angular grid $(\theta_i,\varphi_j)$ is smaller than some prescribed value $\epsilon_h$.

\end{enumerate}

%%%%%%%%%%%%%%%%%%%%%%%%%%%%%%%%
\section{Tests}
\label{sec:tests}
%%%%%%%%%%%%%%%%%%%%%%%%%%%%%%%%

\subsection{Kerr-Schild data}

As a first test of the apparent horizon finder, we use a single black hole 
in Kerr-Schild coordinates (see, e.g., \cite{mtw}) to study its convergence 
properties and robustness. 
Let $M$ and $a$ denote respectively the mass and spin parameter of the black hole. 
In the standard spherical coordinates $(r,\theta,\varphi)$, the polar and equatorial 
coordinate radii of the apparent horizon are given by 
\begin{equation}
r_{\rm po} = \overline{r} , \ \ r_{\rm eq} = \sqrt{ \overline{r}^2 + a^2 } , 
\end{equation}
where $\overline{r} = M + \sqrt{M^2 - a^2}$. The area is given by 
\begin{equation}
A = 4\pi\left( \overline{r}^2 + a^2 \right) . 
\label{eq:kerr_area}
\end{equation}

We first test the convergence property of the code with respect to 
increasing number of collocation points from runs with a black hole of $M=1$ and $a=0.9$. 
The polar radius of the apparent horizon is $r_{\rm po}\approx 1.436$ and the 
equatorial radius is $r_{\rm eq}\approx 1.695$. The analytic data 
$(\gamma_{\hat i\hat j},K_{\hat i\hat j})$
are set on the numerical grid points of the computational domain ranging from 
$r=1$ to $r=5$, which is covered by three spectral domains. The boundary between 
the first and the second domains is at $r=r_{12}=1.5$, whereas that between the 
second and the third domains is at $r=r_{23}=2.5$. In each domain, we use 
$(N_r,N_\theta,N_\varphi)$ collocation points. We also enforce a symmetry with 
respect to the equatorial plane. 
The initial guess for $h$ is a sphere at $r=3$. 

\begin{table}
\caption{Convergence test for a Kerr-Schild black hole with $M=1$ and $a=0.9$. 
Listed are the number of radial collocation points in each domain $N_r$, the 
fractional errors in the polar (equatorial) coordinate radius 
$\Delta r_{\rm po}/r_{\rm po}$ ($\Delta r_{\rm eq}/r_{\rm eq}$) 
and area $\Delta A/A$, the maximum remaining error of 
the expansion function $\Delta \Theta_{\rm max}$ on the horizon and the run times.
We use $N_\theta=(N_r+1)/2$ points in the polar direction and $N_\varphi=1$ in the 
azimuthal direction. We set the iteration parameter $\epsilon_h=10^{-10}$.}
\label{tab:kerr}
\begin{indented}
\item[] \begin{tabular}{c c c c c c}
\hline $N_r$ & $\Delta r_{\rm po}/r_{\rm po}$ & $\Delta r_{\rm eq}/r_{\rm eq}$ &
$\Delta A/A$ & $\Delta\Theta_{\rm max}$ & Time (s) \\
\hline
13 & $2.360\times 10^{-5}$ & $8.003\times 10^{-6}$ & $1.276\times 10^{-5}$ & $4.240\times 10^{-3}$ & 0.507 \\
17 & $1.693\times 10^{-6}$ & $9.705\times 10^{-8}$  & $1.605\times 10^{-6}$ & $5.333\times 10^{-4}$ & 0.747 \\
21 & $1.580\times 10^{-7}$ & $2.145\times 10^{-8}$ & $1.788\times 10^{-7}$ & $6.574\times 10^{-5}$ & 1.129 \\
25 & $1.692\times 10^{-8}$ & $4.235\times 10^{-9}$ & $2.045\times 10^{-8}$ & $8.033\times 10^{-6}$ & 1.615 \\
33 & $1.067\times 10^{-10}$ & $1.707\times 10^{-10}$ & $4.733\times 10^{-10}$ & $1.559\times 10^{-7}$ & 3.059 \\
37 & $1.649\times 10^{-10}$ & $1.325\times 10^{-10}$ & $2.863\times 10^{-10}$ & $2.383\times 10^{-8}$ & 4.286 \\
41 & $1.590\times 10^{-10}$ & $1.053\times 10^{-10}$ & $2.154\times 10^{-10}$ & $3.790\times 10^{-9}$ & 5.722 \\
\hline
\end{tabular}
\end{indented}
\end{table}

Table~\ref{tab:kerr} shows the results 
for increasing $N_r$, with $N_\theta=(N_r+1)/2$ and $N_\varphi=1$. 
We choose the iteration parameter $\epsilon_h=10^{-10}$ in this test 
(see section~\ref{sec:numerical}). In the table, for 
each $N_r$, we list the fractional errors in the polar (equatorial) coordinate 
radius $\Delta r_{\rm po}/r_{\rm po}$ ($\Delta r_{\rm eq}/ r_{\rm eq}$) and 
area $\Delta A/A$, the maximum remaining error in the expansion function 
$\Delta \Theta_{\rm max}$ on the horizon's surface, and the run 
times\footnote{The run times correspond to the CPU 
time the code took to locate the apparent horizon on a 2 GHz Intel Core Duo 
processor. The best-fitted curve suggests that the scaling of the run time is close 
to $N_\theta^2$, which comes from the computation of discrete Legendre transforms.}. 
The error in the area is defined by 
$\Delta A/A := | (A_{\rm ana} - A_{\rm num} )/A_{\rm ana} |$, where the analytic 
result $A_{\rm ana}$ is given by equation~(\ref{eq:kerr_area}) and the numerical 
result is calculated by the integral 
\begin{equation}
A_{\rm num} = \int_S \sqrt{ \hat{q} } h^2 \sin\theta d\theta d\varphi , 
\end{equation}
with $\hat{q}$ being the determinant of the 2-metric on the apparent-horizon's  
surface (expanded onto the basis $\{ {\bf e}_{\hat i} \}$). Explicitly, in terms 
of a general 3-metric $\gamma_{\hat i\hat j}$, $A_{\rm num}$ is given by 
\begin{eqnarray}
\fl
A_{\rm num} = \int_0^{2\pi} \int_0^{\pi} \left[ \left( \gamma_{\hat r\hat r}
h_{,\theta}^{\ 2} + 2\gamma_{\hat r\hat\theta}h h_{,\theta} + 
\gamma_{\hat\theta\hat\theta} h^2 \right) 
\left( \gamma_{\hat r\hat r} h_{,\varphi}^{\ 2} + 
2\gamma_{\hat r\hat\varphi} h h_{,\varphi} \sin\theta  
+ \gamma_{\hat\varphi\hat\varphi} h^2 \sin^2\theta \right) \right. \nonumber \\ 
\left. 
- \left( \gamma_{\hat r\hat r}h_{,\theta}h_{,\varphi}
+ \gamma_{\hat r\hat\theta} h h_{,\varphi} + 
\gamma_{\hat r\hat\varphi} h h_{,\theta} \sin\theta 
+ \gamma_{\hat\theta\hat\varphi} h^2 \sin\theta \right)^2 \right]^{1/2} 
d\theta d\varphi . 
\end{eqnarray}

\begin{figure}
\centering
\includegraphics*[width=8cm]{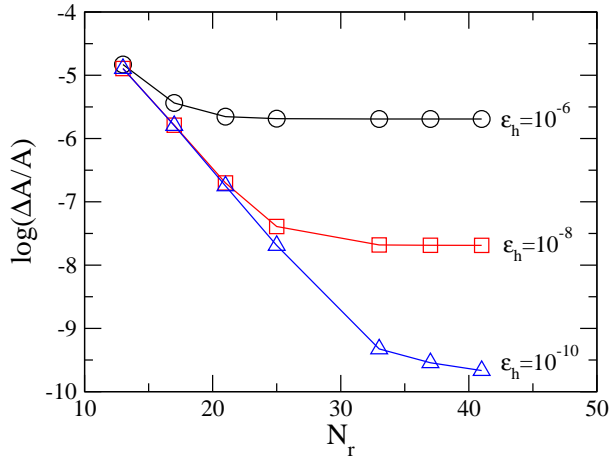}
\caption{Convergence towards zero of the fractional error in the area $\Delta A/A$ 
with the number of collocation points for three different choices of the iteration 
parameter $\epsilon_h$.}
\label{fig:kerr_DeltaA}
\end{figure}

In figure~\ref{fig:kerr_DeltaA} we plot $\Delta A/A$ against $N_r$ to  
explicitly show the convergence behaviour of the finder for three different 
choices of $\epsilon_h$. It can be seen that the error $\Delta A/A$ converges 
exponentially towards zero with the number of points, as expected for spectral 
methods, until the accuracy is limited by the choice of $\epsilon_h$. 
Furthermore, we also see that the number of iterations to a given error level 
$\epsilon_h$ is essentially independent of the value of $\ell_{\rm max}$ used in 
equation~(\ref{eq:h_expand}). This agrees with the conclusions obtained from the 
original Nakamura {\it et al}'s algorithm (or its modifications) as investigated by 
Kemball and Bishop \cite{kem91}.

Next we test the robustness of our finder by performing runs with 
different initial guesses for $h$. We use the same black hole as above 
$(M=1,a=0.9)$, but with a larger computational domain ranging from 
$r=1$ to $r=10$. The boundaries between the different spectral domains 
are $r_{12}=2.5$ and $r_{23}=5.5$. In general, we set up an initial guess 
for $h$ to be the surface of an ellipsoid given by 
\begin{equation}
{x^2\over a^2} + {y^2\over b^2} + {z^2\over c^2} = 1 , 
\label{eq:ellp}
\end{equation}
where $(x,y,z)$ are the Cartesian coordinates relating to the spherical 
coordinates $(r,\theta,\varphi)$ in the standard way. The constants $(a,b,c)$ 
are freely chosen. The initial guess used for the results listed in 
table~\ref{tab:kerr} corresponds to $a=b=c=3$. 
Table~\ref{tab:robust} contains the results for five different initial guesses. 
Case A corresponds to a sphere with a coordinate radius which is about five times 
away from the horizon's surface. On the other hand, the initial surface for case
B is a sphere located entirely inside the apparent horizon. The initial guess for 
case C is an ellipsoid enclosing the horizon. Finally, cases D and E represent 
initial surfaces which cross the horizon. The results show that our finder can 
locate the apparent horizon to the same accuracy with all the five (quite generic) 
choices of $(a,b,c)$.

\begin{table}
\caption{Robustness test for the same black hole used in table~\ref{tab:kerr}.
The polar coordinate radius of the apparent horizon is $r_{\rm po}\approx 1.436$ 
and the equatorial coordinate radius is $r_{\rm eq}\approx 1.695$. The initial 
guess for the 2-surface $h$ is given by the surface of an ellipsoid with axes 
$(a,b,c)$ defined in equation~(\ref{eq:ellp}). We use collocation points $N_r=25$ 
and $N_\theta=13$ for all the five cases considered, but $N_\varphi=1\ (4)$ for 
cases A and B (C-E).}
\label{tab:robust}
\begin{indented}
\item[] \begin{tabular}{c c c}
\hline Case & $(a,b,c)$ & $\Delta A/A \ (10^{-6})$ \\
\hline
A & $(8,8,8)$ & $4.04871$ \\
B & $(1.2,1.2,1.2)$ & $4.04905$ \\
C & $(4,6,8)$ & $4.04875$ \\
D & $(2,3,1.2)$ & $4.04910$ \\
E & $(1.2,1.5,2)$ & $4.04909$ \\
\hline
\end{tabular}
\end{indented}
\end{table}

One of the main requirements of an apparent horizon finder is speed. This is 
in particular an important issue if the finder has to run frequently during a 
simulation. 
In order to compare the speed of our finder with some other commonly used 
methods, we take the data given by Schnetter \cite{sch03}. 

In table 5 of \cite{sch03} Schnetter compared the run times to locate the 
apparent horizon of a Kerr-Schild black hole with $M=1$ and $a=0.6$ for 
his elliptic method and two other methods, namely the fast-flow 
\cite{gun98} and minimization \cite{ann98} algorithms.  
The fastest case (0.5 s on a 1.2 GHz processor) was obtained by his 
elliptic method with the initial guess being a sphere at $r=2$. 
The error in the area is $\Delta A/A = 9\times 10^{-3}$ (according 
to table 4 of \cite{sch03}). 
For comparison, the fast-flow and minimization algorithms took more than 10 s and 
90 s respectively in the test \cite{sch03}. 

We have performed tests with the same black hole and initial 
guess, and found that our finder took 0.129 s (on our 2 GHz processor) to locate 
the horizon to the accuracy $\Delta A/A = 2\times 10^{-5}$ using the resolution 
$(N_r,N_\theta,N_\varphi)=(7,5,1)$ with $\epsilon_h=10^{-8}$.
We have also used Thornburg's finder \program{AHFinderDirect} \cite{tho04} (which is 
implemented within the \program{CACTUS} computational toolkit \cite{cactus}) to perform 
the same test using Cartesian grid resolutions $(N_x,N_y,N_z)=(31,31,19)$ with 
$\Delta x=\Delta y=\Delta z=0.2$ in bitant symmetry. We found that his finder took 
1.004 s (on our 2 GHz processor) to locate the apparent horizon to the accuracy 
$\Delta A/A = 3\times 10^{-4}$.

We note that the above test does not represent a direct comparison between the 
different algorithms because of the different grid structures (Cartesian versus spherical coordinates), code implementations, memory usage and computer systems. 
Nevertheless, we can conclude that for this particular test, to obtain about the same 
accuracy level, our spectral-method-based finder is as efficient as the finders 
developed by Schnetter \cite{sch03} and Thornburg \cite{tho04}.

\subsection{Brill-Lindquist data} 

\begin{table}
\caption{Schwarzschild black hole offset from the coordinate origin. The hole is located at 
the Cartesian coordinates $(d/\sqrt{2},d/\sqrt{2},0)$. Listed are the offset $d$, the fractional 
error in the area $\Delta A/ A$, and the maximum remaining error of the expansion function
on the horizon $\Delta \Theta_{\rm max}$. }
\label{tab:shiftBH}
\begin{indented}
\item[] \begin{tabular}{c c c}
\hline $d$ & $\Delta A/A$ & $\Delta\Theta_{\rm max}$ \\
\hline
$0.1$ & $9\times 10^{-6}$ & $3\times 10^{-3}$ \\
$0.2$ & $3\times 10^{-6}$ & $5\times 10^{-3}$ \\
$0.3$ & $2\times 10^{-6}$ & $2\times 10^{-2}$ \\
$0.4$ & $1\times 10^{-4}$ & $5\times 10^{-2}$ \\
\hline
\end{tabular}
\end{indented}
\end{table}

In this part, we test our finder using the Brill-Lindquist data \cite{bri63}. 
This is a classic test involving multiple black holes used in numerical relativity. 
The 3-metric is conformally flat, $\gamma_{ij} = \phi^4 f_{ij}$, and is 
time symmetric (i.e. $K_{ij}=0$). For two black holes, $\phi$ is given by 
\begin{equation}
\phi = 1 + {M_1\over 2| \vec{r} - \vec{r}_1 | } + { M_2\over 2| \vec{r} - \vec{r}_2 | } , 
\end{equation}
where $M_i$ ($i=1,2$) is the mass of the $i$th black hole and $\vec{r}_i$ are the 
coordinate positions of the holes. 

We first begin with a single black hole ($M_2=0$), in which case the problem
is equivalent to a Schwarzschild black hole in isotropic coordinates offset
from the coordinate origin. The apparent horizon is a coordinate sphere of
radius $M_1/2$ with respect to the center of the hole. The area of the horizon
is $A=16\pi M_1^2$. We set $M_1=1$ and the coordinate position of the hole at
$\vec{r}_1=(x_1,y_1,z_1)=(d/\sqrt{2}, d/\sqrt{2}, 0)$. We have varied $d$ in
order to verify that our finder also works when the center of the spherical
harmonics is offset from the center of the horizon. Table~\ref{tab:shiftBH}
lists the results for four different values of $d$. The initial guesses are
always $a=b=c=1$ in equation~(\ref{eq:ellp}).  We use three spectral domains
to cover the spatial slice up to $r=1.5$, with collocation points
$(N_r,N_\theta,N_\varphi)=(33,17,16)$ in each domain.  The boundaries between
the domains are $r_{12}=0.5$ and $r_{23}=0.8$.  Similar to \cite{gun98,kem91},
we see that the accuracy drops quite significantly for very distorted surfaces
with respect to the coordinate origin. In particular, the error in the area
$\Delta A/A$ increases by almost two orders of magnitude when $d$ increases
from 0.3 to 0.4; this error could be reduced using higher grid 
resolution\footnote{The error $\Delta A/A$ drops down to $2\times 10^{-5}$ for 
the case $d=0.4$ using collocation points $(N_r,N_\theta,N_\varphi)=(33,25,24)$.}.  
Nevertheless, it is worth to point out that the original
Nakamura {\it et al} spectral algorithm \cite{nak84} would not produce any
results for $d=0.3$ and 0.4 because equation~(\ref{eq:a00}) has no roots
\cite{kem91}. We also see that the results are essentially independent of the
direction of the offset.

Next we turn to Brill-Lindquist data for two black holes of equal mass. In particular, we 
take $M_1=M_2=1$ in the test. 
The data forms a one-parameter family parameterized by the coordinate separation $d$ 
between the holes. When they are far apart, each hole has an individual apparent horizon. For 
small separation, there is a single common apparent horizon enclosing both holes. Determining 
the critical separation at which the common horizon appears in this two black-hole spacetime 
is a standard test problem for apparent horizon finders. 
The critical separation obtained originally by Brill and Lindquist is 
$d_c=1.56$ \cite{bri63}, while more recent results suggest that $d_c\approx 1.53$ 
(e.g., \cite{tho04,kem91,alc00,sho00}). 
In particular, we note that Thornburg \cite{tho04} and Shoemaker {\it et al} \cite{sho00} report 
very close results at $d_c=1.532$ and $d_c=1.535$ respectively. 
Nevertheless, Thornburg reports $A=196.407$ for the area of the critical apparent horizon, which 
is quite different from the value $A=184.16$ obtained by Shoemaker {\it el al.}.  

\begin{figure}
\centering
\includegraphics*[width=8cm]{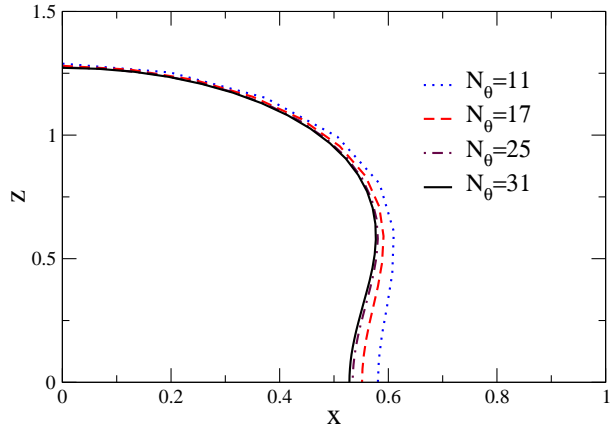}
\caption{Position of the common apparent horizon on the $x$-$z$ plane for 
Brill-Lindquist data with $d=1.532$. The black holes are centered at $z=\pm d/2$ 
along the $z$-axis. 
The results obtained by four different 
values of $N_\theta$ are shown (with $N_r=41$ and $N_\varphi=1$ fixed). }
\label{fig:brillBH}
\end{figure}

Here we test our finder by trying to find a common horizon at the critical separations, as 
reported by Thornburg \cite{tho04} and Shoemaker {\it et al.} \cite{sho00}. 
The black holes are on the $z$-axis, with their centers 
at $z= \pm d/2$. In the test, we use four spectral domains to cover the spatial slice up to 
$r=2$. The boundaries between the domains are $r_{12}=0.5$, $r_{23}=1$ and $r_{34}=1.5$. 
The initial guesses are $a=b=c=2$ in equation~(\ref{eq:ellp}). We use 
$(N_r,N_\theta,N_\varphi)=(41,31,1)$ in each domain and the iteration parameter 
$\epsilon_h=10^{-6}$. 
Our finder reports a common horizon at $d=1.532$ (Thornburg's critical 
value) with the area of that horizon 
determined to be $A=196.417$, which agrees to Thornburg's value to $0.005\%$. 
The maximum remaining error of the expansion function on the horizon is 
$\Delta \Theta_{\rm max}=7\times 10^{-4}$. The finder took 59.2 s to locate the horizon.
We note that increasing $\epsilon_h$ to $10^{-4}$ would reduce the run time to 23.3 s, without
changing the three significant figures of $A$. 
On the other hand, for the same grid setting and parameters, our finder does not find a 
common horizon at the critical value $d=1.535$ reported by Shoemaker {\it et al.}.
In general, for $d > 1.532$, we find two disjoint apparent horizons surrounding $\vec{r}_1$ 
and $\vec{r}_2$ by setting the coordinate origin for the apparent horizon finder
(see section~\ref{sec:define}) separately at around the points $\vec{r}_1$ and $\vec{r}_2$. 
In figure~\ref{fig:brillBH} we show the position of the common apparent 
horizon on the $x$-$z$ plane for the case $d=1.532$. The results obtained by four different 
values of $N_\theta$ (with $N_r=41$ and $N_\varphi=1$ fixed) are plotted together to show the 
convergence of the horizon.

\section{Conclusions}
\label{sec:conclude}

In this paper we have presented a new apparent horizon 
finder which is based on spectral methods. Our proposed algorithm does not 
need to treat the $\ell=0$ coefficient of the spherical harmonics decomposition 
separately as required in previous spectral apparent horizon finders 
\cite{nak84,kem91}. Hence, this leads to a faster and more robust 
finder based solely on spectral methods.
We have made a performance comparisons with other apparent 
horizon finders using the Kerr-Schild data. Our finder is much faster (by 
orders of magnitude) than other commonly used methods (e.g., the fast-flow and
minimization algorithms). It is also as efficient as the currently fastest 
methods developed recently by Schnetter \cite{sch03} and 
Thornburg \cite{tho04}. We have also shown that our finder is capable of 
locating the horizon of a shifted Schwarzschild black hole with a large offset 
from the coordinate origin. This would not be possible by using the 
original Nakamura {\it et al} spectral algorithm \cite{nak84}
because equation~(\ref{eq:a00}) has no roots if the offset is too large. 
We have also tested our finder for a two black-hole spacetime using the 
Brill-Lindquist data. In particular, we have verified previous results 
on the critical separation at which a common horizon appears in this 
spacetime. 

Our apparent horizon finder is implemented within the C++ library LORENE for
numerical relativity \cite{lorene}, and is freely available. The finder should
be easily adopted in spectral-method-based evolution codes
\cite{bon04,sch04,tic06,boy07}, particularly to those using shell-like domains
in spherical coordinates. Comparing to other freely available apparent horizon
finders which are based on the finite-differencing method (\program{AHFinder}
\cite{alc00} and \program{AHFinderDirect} \cite{tho04}), our finder also
represents another option available to finite-differencing evolution codes,
with some interpolation to be implemented between the finite-difference and
spectral grids as, for example, in \cite{dim05}.

\vspace{0.2in}
\noindent
{\it Note added:} After we have submitted the paper, Tsokaros and Ury\={u} 
informed us that they had recently developed an apparent horizon finder based 
on the same formulation presented in this paper \cite{tso07}. 
Their work and ours were done independently.

\vspace{0.2in}
\noindent
{\bf Acknowledgments}
\vspace{0.1in}

\noindent
We thank Jos\'{e} Luis Jaramillo for very helpful discussions. We also 
thank Antonios Tsokaros and K\={o}ji Ury\={u} for bringing their work 
\cite{tso07} to our attention. 
LML is supported in part by the Hong Kong Research Grants Council 
(Grant No: 401905 and 401807) and a postdoctoral fellow scheme at the Chinese 
University of Hong Kong. JN was supported by the A.N.R. grant
06-2-134423 entitled ``Mathematical methods in general relativity'' (MATH-GR).

\vspace{0.2in}
\noindent
{\bf References}
\vspace{0.1in}

%%%%%%%%%%%%%%%%%%%%%%%%%%%%%%%%%%%%%% 
%% reference 
%%%%%%%%%%%%%%%%%%%%%%%%%%%%%%%%%%%%%% 
\bibliographystyle{prsty}

\end{document}